\documentclass[preprint,5p,number,sort&compress]{elsarticle}

\usepackage{dcolumn}% Align table columns on decimal point
\usepackage{bm}% bold math
\usepackage{textcomp}
\usepackage{upgreek}
\usepackage{graphicx}
\usepackage{hyperref}% add hypertext capabilities
\usepackage[load-configurations=abbreviations,detect-all=true]{siunitx}
\usepackage[version=3]{mhchem}
\usepackage[mathlines]{lineno}% Enable numbering of text and display math
\usepackage[usenames,dvipsnames,svgnames,table]{xcolor} % color text
\usepackage[normalem]{ulem} % strikeout
\usepackage{transparent}
\usepackage{amsmath}

\usepackage{xspace}

\newcommand{\TMIIm}{\mbox{\emph{Topmetal-II\raise0.5ex\hbox{-}}}\xspace}

\newcommand{\Ar}{\ensuremath{^{39}\text{Ar }}}
\newcommand{\Cf}{\ensuremath{C_\text{f}}\xspace}
\newcommand{\Cs}{\ensuremath{C_\text{s}}\xspace}
\newcommand{\Cin}{\ensuremath{C_\text{in}}\xspace}
\newcommand{\Rin}{\ensuremath{R_\text{in}}\xspace}
\newcommand{\Vin}{\ensuremath{V_\text{in}}\xspace}
\newcommand{\Vout}{\ensuremath{V_\text{out}}\xspace}
\newcommand{\dQ}{\ensuremath{\Delta Q}\xspace}
\newcommand{\dt}{\ensuremath{\Delta t}\xspace}
\newcommand{\D}{\ensuremath{\Delta}\xspace}
\newcommand{\Ohm}{\ensuremath{\Omega}\xspace}
\newcommand{\gm}{\ensuremath{g_\text{m}}\xspace}
\newcommand{\Ids}{\ensuremath{I_\text{ds}}\xspace}
\newcommand{\Rds}{\ensuremath{R_\text{ds}}\xspace}
\newcommand{\Vds}{\ensuremath{V_\text{ds}}\xspace}
\newcommand{\us}{\textmu s}
\newcommand{\Vgs}{\ensuremath{V_\text{gs}}\xspace}
\newcommand{\waveforms}{\text{waveforms}}

\newcommand{\CfRds}{\ensuremath{C_\text{f}\cdot R_\text{ds}}\xspace}
\newcommand{\CsRds}{\ensuremath{C_\text{s}\cdot R_\text{ds}}\xspace}

\DeclareSIUnit\keV{keV}
\DeclareSIQualifier\ee{ee}
\DeclareSIQualifier\nr{nr}
\DeclareSIUnit\pe{p.e.}

\journal{\phantom{Nuclear Instruments and Methods in Physics Research A}}

\begin{document}

\begin{frontmatter}

\title{Demonstrating the Q-Pix front-end using discrete OpAmp and CMOS transistors}

%\ead{This paper and the work therin is presented in memory of Gary Varner's pioneering work in Q-Pix}
%\author{Author list to be determined\fnref{fn0}}
\author[lbl]{Peng Miao}
\author[UTA]{Jonathan~Asaadi}
\author[WC]{James~B.~R.~Battat}
\author[lbl]{Mikyung~Han}
\author[UH,UTA]{Kevin~Keefe}
\author[UH]{S.~Kohani}
\author[UTA,IFS]{Austin~D.~McDonald}
\author[UTA]{David~Nygren}
\author[UTA]{Olivia~Seidel}
\author[lbl,UTA]{Yuan~Mei\corref{cor0}}\ead{ymei@lbl.gov}

\address[lbl]{Lawrence Berkeley National Laboratory, Berkeley, CA 94720, USA}
\address[UTA]{Department of Physics, University of Texas at Arlington, Arlington, TX 76019, USA}
\address[IFS]{IF Scientific, Arlington, TX 76018, USA}
\address[WC]{Wellesley College, Physics Department, 106 Central Street, Wellesley, MA 02481}
\address[UH]{Department of Physics and Astronomy, University of Hawaii, Honolulu, HI 96822, USA}

\cortext[cor0]{Corresponding author}
%\fntext[fn0]{Current address: }

\begin{abstract}

Using Commercial Off-The-Shelf (COTS) Operational Amplifiers (OpAmps) and Complementary Metal-Oxide Semiconductor (CMOS) transistors, we present a demonstration of the Q-Pix front-end architecture, a novel readout solution for kiloton-scale Liquid Argon Time Projection Chamber (LArTPC) detectors.  The Q-Pix scheme employs a Charge-Integrate/Reset process based on the Least Action principle, enabling pixel-scale self-triggering charge collection and processing, minimizing energy consumption, and maximizing data compression.  We examine the architecture's sensitivity, linearity, noise, and other features at the circuit board level and draw comparisons to SPICE simulations.  Furthermore, we highlight the resemblance between the Q-Pix front-end and Sigma-Delta modulator, emphasizing that digital data processing techniques for Sigma-Delta can be directly applied to Q-Pix, resulting in enhanced signal-to-noise performance.  These insights will inform the development of Q-Pix front-end designs in integrated circuits (IC) and guide data collection and processing for future large-scale LArTPC detectors in neutrino physics and other high-energy physics experiments.

  % \begin{description}
  % \item[Purpose]
  % \end{description}
\end{abstract}

\end{frontmatter}

\section{Introduction}\label{sec:intro}
%\subsection{Context}
The Deep Underground Neutrino Experiment (DUNE) is the next-generation neutrino experiment aiming to answer fundamental questions about the universe and is aimed to be the U.S's flagship particle physics experiment over the next decade.\cite{DUNE:2016hlj}

The primary detector technology selected by DUNE's Far Detector (FD) is the Liquid Argon Time Projection Chamber (LArTPC)\cite{Nygren:1974nfi,Willis:1974gi,Rubbia:1977zz}.  LArTPCs have been successfully used in particle physics experiments for many years as they offer excellent high-precision 3D imaging of ionization tracks produced by charged particles\cite{Rubbia:2011ft,Anderson:2012vc,MicroBooNE:2016pwy}.  Typically, the readout system for LArTPCs has two or more assembled arrays, or planes, of sense wires. However, such wire-based readout systems can produce ambiguities in data when drifting electrons are isochronous or parallel to a wire orientation\cite{Adams_2020}.  To overcome this issue, the readout system for LArTPCs is evolving from wire-based planes to charge-sensitive pixel arrays, known as pixelated readout systems\cite{Asaadi:2018oxk}.

To ensure that the data extracted from the DUNE FD detectors are of the highest quality and to reveal any subtle new phenomena near the limit of detection, a true 3D pixelated signal capture approach, known as the Q-Pix concept\cite{nygren2018qpix}, has been proposed as a potential readout solution.

Q-Pix captures detailed waveforms of the incoming ionization charge by measuring time per unit charge rather than using the classic continuous sampling and digitization method.  Compared to traditional readout electronics designs, Q-Pix is much simpler yet provides higher charge precision and spatial resolution with a lower threshold.  This makes it ideal for measuring very rare and low energy signals in a LArTPCs, such as those produced by solar neutrinos or supernova explosions, while being capable of recording large ionization signals with high fidelity as seen in GeV-scale neutrino interactions coming from accelerator produced neutrinos.

Previous simulation studies have demonstrated promising results for the Q-Pix pixelization concept\cite{nygren2018qpix,Kubota_2022}; however, to validate its performance and feasibility for future use, it is still essential to conduct a real-world demonstration based on actual hardware.  Such testing offers valuable insights into the practical implementation and reveal potential critical limiting factors, thereby providing opportunities for enabling refinement and improvement of the Q-Pix design.  Successfully demonstrating Q-Pix using COTS hardware will also be a crucial step towards an effective ASIC design for a large-scale pixelated readout system used in the LArTPCs of DUNE.

\section{Q-Pix design and demonstration}\label{sec:design}
The key principle of Q-Pix is the concept of `Least Action', which can be described as the pixel readout remaining in a low-power quiescent state during periods when no ionization charge is present.  This state can be thought of as the pixel being ``OFF''.  However, the pixel readout needs to be ready and capable of collecting charge and transitioning quickly to an ``ON'' state.  The ``ON'' state must still satisfy the low-power requirements placed on the electronics present in a liquid argon environment to minimize heating of the medium.  To accomplish this, a simple Charge-Integrate/Reset (CIR) circuit block with a time-stamping mechanism based on free-running local clocks is employed.  Such a seemingly simple architecture is able to achieve this principle of `least action' in an unorthodox yet surprisingly natural overall solution.

\subsection{Q-Pix reset scheme}
\begin{figure}[!htb]
\centering
\includegraphics[width=0.99\linewidth]{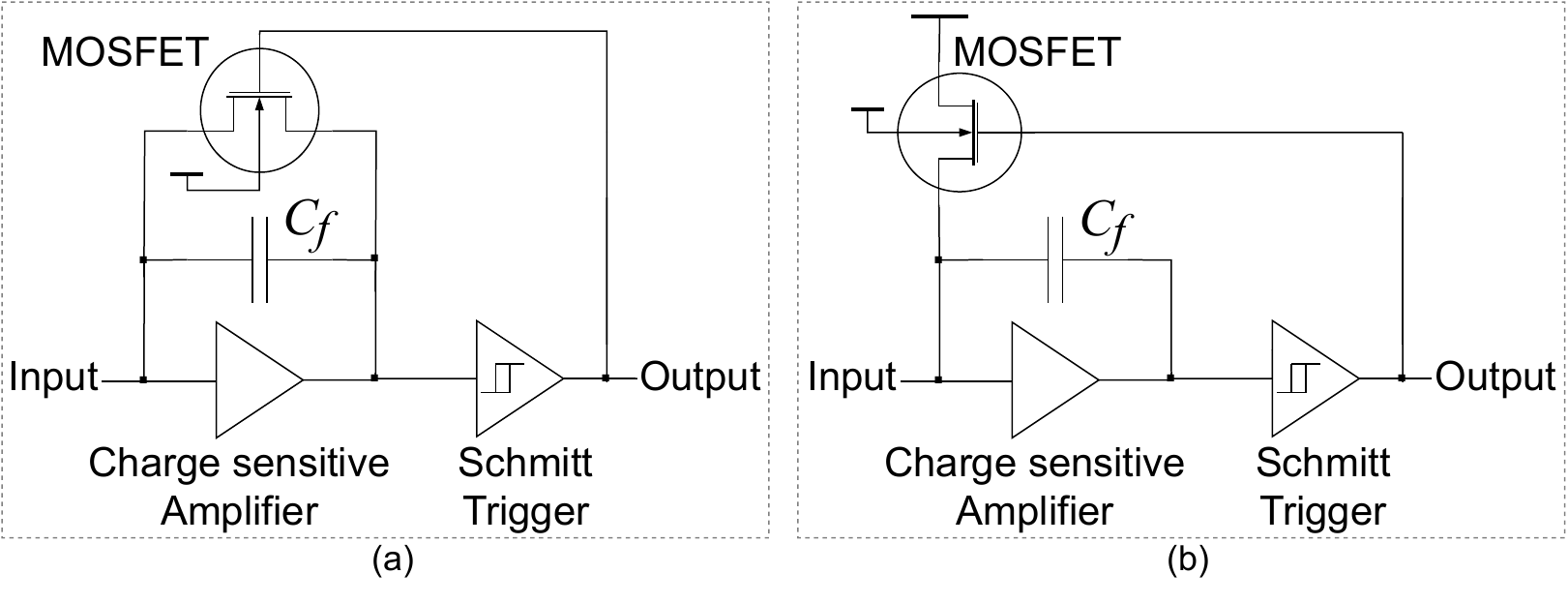}
\caption{\label{fig:schemes} Q-Pix front-end schemes.  Panel (a) depicts the reset scheme.  When the charge integrated by the charge-sensitive amplifier reaches the Schmitt Trigger's threshold, the MOSFET discharges the charge on \Cf, resetting the amplifier and generating a short, standardized reset signal.  Panel (b) illustrates the replenishment scheme.  When the output of the CSA reaches the threshold of the Schmitt Trigger, the MOSFET is activated, replenishing \Cf with a preset current.  This current flows opposite the input current direction from the LArTPC detector.}
\end{figure}

The initially proposed Q-Pix scheme\cite{nygren2018qpix} is shown in Fig.~\ref{fig:schemes}(a).   This scheme uses a charge-sensitive amplifier (CSA) for charge integration.  Once the voltage of the feedback capacitor \Cf reaches the preset threshold, the Schmitt trigger will turn on the MOSFET switch to quickly short \Cf and thus discharging and resetting the Q-Pix readout to a stable (arbitrary) baseline.  This constitutes the ``Charge-Integrate Reset'' (CIR) circuit.  The time at which the reset occurs is recorded using a local clock and represents the data output of the Q-Pix pixel.  This structure is referred to as the ``reset scheme.''

\subsection{Q-Pix replenishment scheme}

An alternative architecture which can accomplish the same goal, with potentially better performance is also considered for Q-Pix, known as the  ``replenishment scheme'' as depicted in Fig.~\ref{fig:schemes} (b).  In this scheme, the charge integration process is the same as before, however, instead the MOSFET switch is replaced by a MOSFET controlled current source.  Whenever the comparator triggers a transition (representing that \Cf has reached the desired threshold voltage), the MOSFET replenishes a charge of $ \Delta Q =I\cdot\Delta t $, where \dt is the pulse width.  This constitutes the ``Charge-Integrate Replenishment'' (CIR) circuit.  The \dQ then forms a charge quantum of each measurement and the time at which the replenishment is recorded using a local clock.

\subsection{Implementation of the CIR circuit}
When deciding between the ``reset'' and ``replenishment'' architecture a number of factors must be considered.  In the ``reset'' scheme, the MOSFET switch ideally operates in either open (disconnected) or closed (conducting) state.  When the amplifier is integrating charge, the switch must be in a fully open state to prevent any charge loss.  This requires negligible leakage charge passing through the switch compared to the input charge.  During the reset of \Cf, the switch should be fully closed.  This means that the current through the MOSFET must be sufficiently large to ensure that the \Cf can be reset to the baseline level as quickly as possible.  It should be noted, that the reset circuit as outlined above will suffer from signal charge loss during the reset.  In order to prevent this charge loss, and the subsequently nonlinear errors that would arise from this, an additional switch would be required to disconnect the input to the CSA during the reset.  This may be fine for highly integrated ASICs with minimal noise as the reset process can be controlled to be brief enough to ensure that the leaked input charge during the MOSFET's ON time is negligible.  However, discrete-component designs with more uncertain parasitic parameters and greater susceptibility to environmental noise may face many difficulties.

These problems are largely absent for the ``replenishment'' architecture as both the leakage current from the MOSFET current source can be neglected (or calibrated away) as well as an absence of charge loss during the CIR process.  Even if there is a significant input charge during the discharging process, as long as the MOSFET current and the discharge time are fixed, the quantization \dQ of each measurement will remain constant.  This allows the Q-Pix to discharge the \Cf through the MOSFET and integrate with the input current simultaneously without loss of charge.  The replenishment scheme also brings additional advantages.  The voltage on the Source and Drain of the MOSFET remain constant throughout the entire CIR process, making it possible to precisely control the MOSFET current and thus easily fine-tune the desired \dQ.  Furthermore, it minimizes the potential influence of the MOSFET parasitic parameters on the CSA feedback loop.  For the above reasons, the replenishment scheme was selected for implementation in this work, as described below.

\subsection{Q-Pix replenishment front-end with clocked reset}
\begin{figure}[!htb]
\centering
\includegraphics[width=0.99\linewidth]{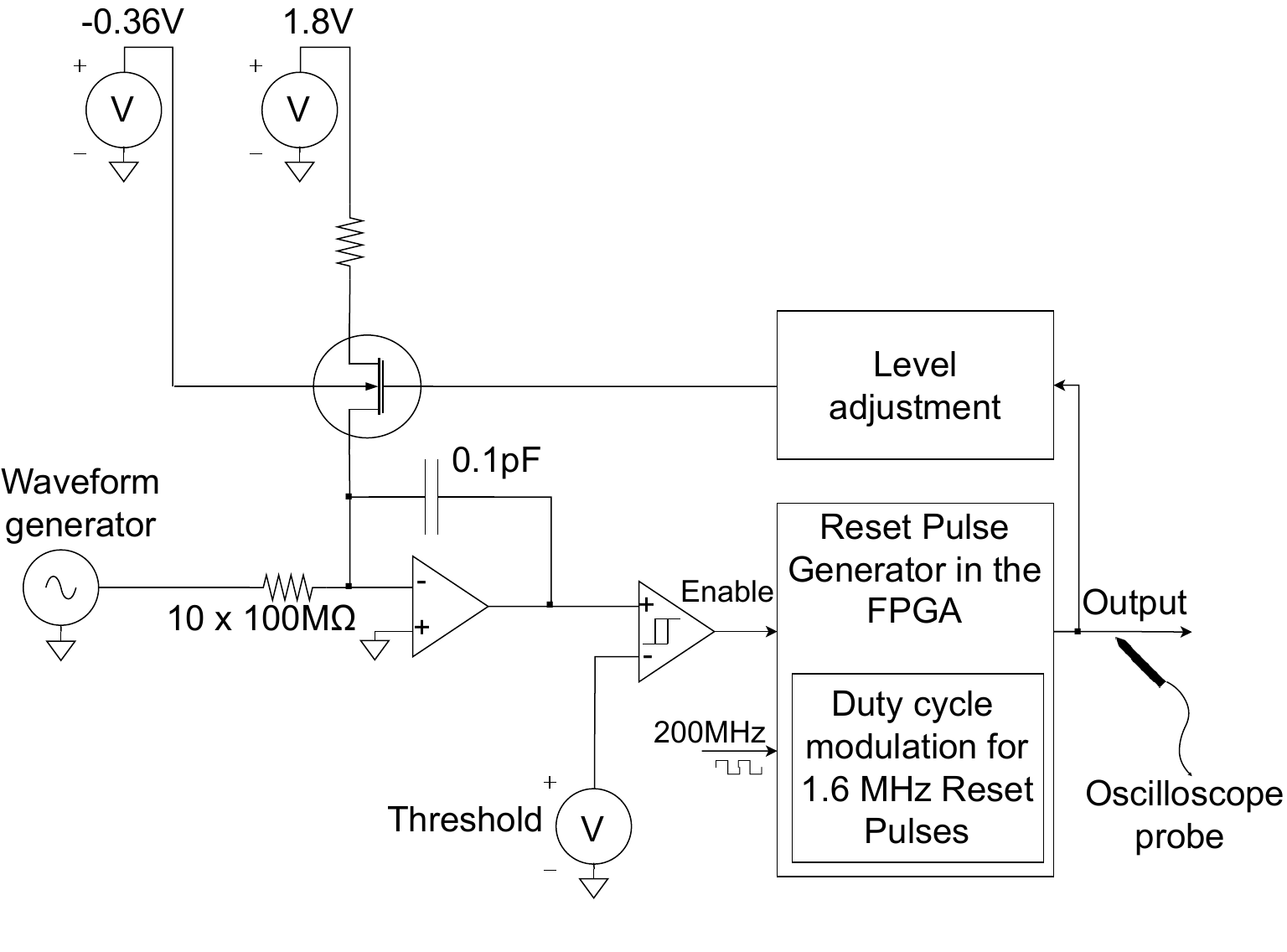}
\caption{\label{fig:frontend} Q-Pix front-end with clocked reset.  The simplified Q-Pix front-end demonstration board schematic is presented, which employs the replenishment scheme.  Due to environmental noise pickup and parasitic parameters, maintaining consistent \dQ across measurements was challenging.  This issue was resolved by utilizing a D-flip-flop to sample the output of the Schmitt trigger, ensuring a consistent and appropriate reset pulse width sent to the MOSFET Gate every time the Schmitt trigger is activated.  This resulted in a clocked reset and a minimum effective pulse width for each reset, controlled by duty cycle modulation implemented on an FPGA.}
\end{figure}

Fig.~\ref{fig:frontend} represents a simplified schematic of the Q-Pix front-end demonstration board which was developed for this work.  To achieve a small charge quantum \dQ for high charge resolution, the capacitance value of \Cf was chosen to be as small as possible.  The smallest readily available COTS discrete capacitor has a capacitance value of \SI{0.1}{pF}.  However, the parasitic capacitance of the CSA and other discrete components is typically within the same order of magnitude, if not larger.  Consequently, the system is highly susceptible to the parasitic parameters of the discrete components and the Printed Circuit Board (PCB) itself.

%In the initial Q-Pix design, the Schmitt trigger is directly connected to the Gate of the MOSFET, and each output `1' from the Schmitt trigger represents a single measurement of one \dQ.  This digital measurement result would then be time-stamped using a local clock to achieve a high level of compression for the detector data, which may only contain a very low rate of intrinsically interesting neutrino events.  This is a straightforward and effective reset generation approach.

During the first attempt at implementing the replenishment scheme the MOSFET was directly controlled by the output of the Schmitt trigger.  However, it was found during testing that maintaining a consistent \dQ was difficult, especially when \dQ was tuned to $\sim \mathcal{O}($fC) or smaller.  For a \SI{0.1}{pF} feedback capacitor, achieving a \SI{1}{fC} \dQ implies a threshold difference of \SI{10}{mV}.  This small threshold voltage was found to be highly susceptible to interference from system noise, leading to random fluctuations in \dQ in the discrete component implementation.  Moreover, the approach of coupling the MOSFET directly to the output of the Schmitt trigger also limited the achievable minimum \dQ.  This is owed to the fact that \dQ depends on the MOSFET current and the reset pulse width.  The width of the reset pulse is determined by the difference between the high and low threshold values of the Schmitt trigger as well as the response speed of the entire feedback loop.  This response speed affects the time it takes for the CSA's output to recover from the high threshold to the low threshold under the same discharge current.  Although increasing the MOSFET current can shorten the recovery time, the \dQ equals to the product of the MOSFET current and the reset pulse width.

The straightforward solution used in this implementation is to force a fixed and appropriate reset pulse width sent to the MOSFET Gate every time the Schmitt trigger is activated.  This is achieved employing a simple D-flip-flop (DFF) to sample the output of the Schmitt trigger, resulting in a clocked reset.  This is the purpose of the Reset Pulse Generator shown in Fig.~\ref{fig:frontend}, where the duty cycle modulation controls each reset's minimum effective pulse width.  We utilize an FPGA to implement this function, with an equivalent reset sampling rate of \SI{1.6}{MHz}.  The upper limit of this frequency is primarily determined by the response speed of the entire feedback loop formed by the CSA, Schmitt trigger, and MOSFET.  The input clock for the FPGA operates at a higher speed of \SI{200}{MHz}, which facilitates precise adjustment of the reset pulse's duty cycle.  The reason for adjusting the duty cycle will be further explained in the following sections. %Parasitic capacitance in the feedback loop also contributes to adverse effects.

\subsection{Expected \dQ range and sampling rate}

The reset pulses from the Schmitt trigger in the replenishment scheme with the clocked reset are still the observed digital outputs to be further processed online or sent to a potential back-end data acquisition system for offline analysis.  After reconstruction, the input current waveform information from the LArTPC detectors remains preserved.

The accuracy of this reconstruction for a LArTPC detector is limited by diffusion of the drifting electrons, up to approximately \SI{4}{mm} RMS transverse and \SI{14}{mm} RMS longitudinal at the maximum drift distance of around \SI{7}{\meter}.  Additionally, track-pair separation and practical minimum signal size suggest a pixelization scale of $4 \times 4\,\si{\milli\meter\squared}$, which is likely near a soft optimum.  For a $4 \times 4\,\si{\milli\meter\squared}$ pixel, \dQ might plausibly correspond to about $1/5$ of a track crossing the pixel centrally, which is approximately \SI{1}{mm}\cite{nygren2018qpix}.  The charge generated by a \SI{1}{mm} track segment for a Min-I particle in LAr is around \si{5000} electrons\cite{Dwyer:2018phu}.  Therefore, taking \SI{1}{mm} as the sample size suggests \dQ should be around \si{5000} electrons or $5/6$\,\si{\femto\coulomb}.  Simulation results of the Q-Pix circuit with simulated signals from neutrino interactions also suggest that an optimum value for \dQ lies between 0.3 and \SI{1.0}{fC}\cite{nygren2018qpix}.  This range is the design target for the CMOS implementation of the Q-Pix frontend currently under investigation and helped set the target for this discrete component implementation.

In the replenishment scheme, the quantization \dQ is proportional to the MOSFET's Source-Drain current.  A ``Level Adjustment circuit'' must be employed to independently adjust the precise high and low levels of the reset pulse before sending it to the MOSFET's Gate.  By selecting an appropriate high level, the Source-Drain current of the MOSFET can be adjusted to achieve the desired value of \dQ.  Conversely, a sufficiently low Gate level can minimize the leakage current of the MOSFET in the off state.

The effective sampling rate of Q-Pix is another key parameter to consider.  On the one hand, for a Min-I track segment that arrives normally and assumes an electron drift velocity in LAr of \SI{1.6}{mm\per\micro\second}\cite{Walkowiak:driftV}, a ``waveform''
sample would be taken on average every \SI{0.63}{\us}.  On the other hand, in practice, the bandwidth of the entire feedback loop sets the upper limit of the achievable sampling rate.  For the discrete component Q-Pix demonstration, a maximum effective sampling rate of approximately \SI{1.6}{MHz} (\SI{0.625}{\micro\second}) is sufficient.

\subsection{Setup for characterization}

\begin{figure}[!htb]
\centering
\includegraphics[width=0.95\linewidth]{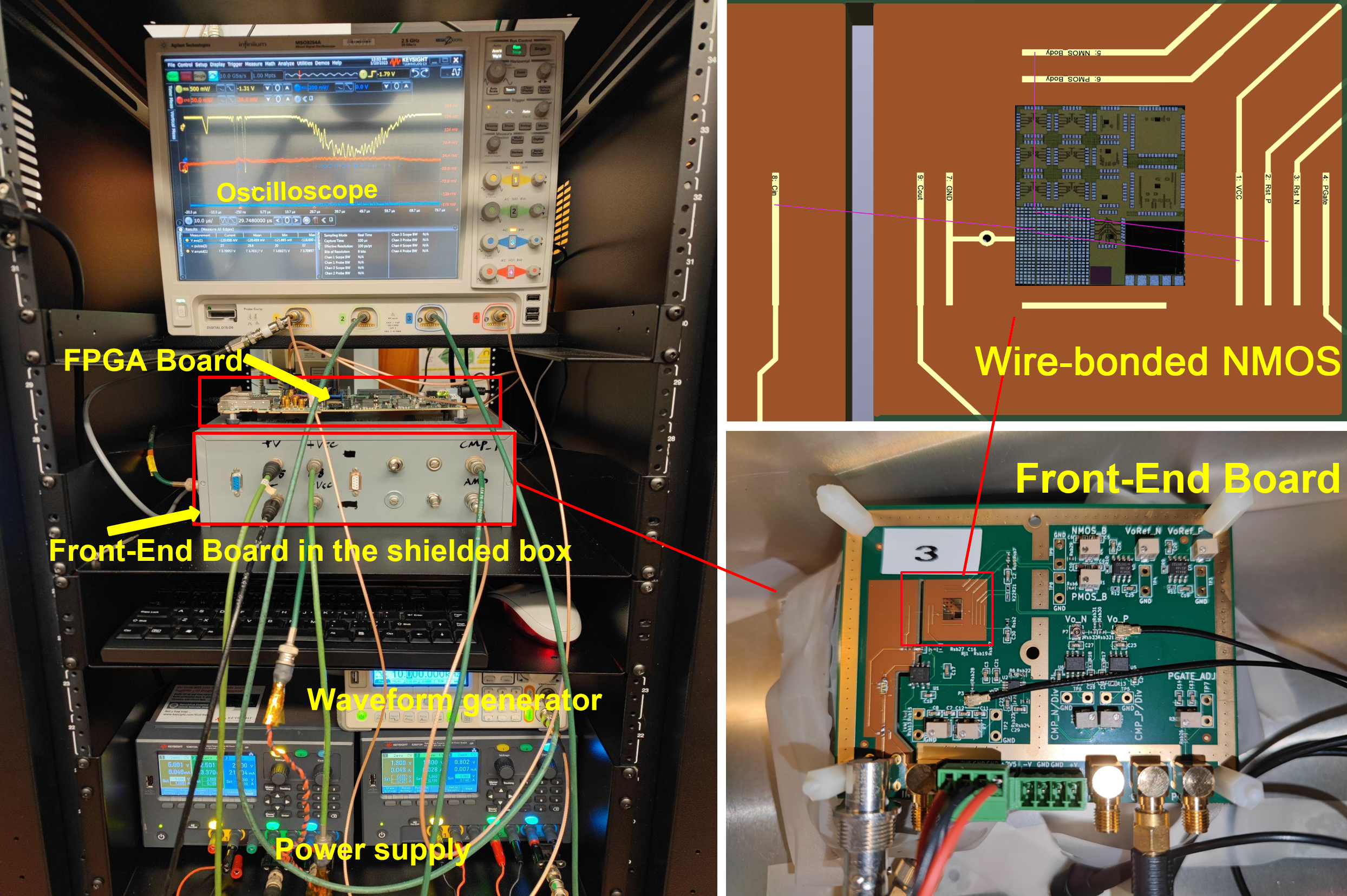}
\caption{\label{fig:setup} Setup for characterization.  This illustration demonstrates the comprehensive setup for Q-Pix characterization.  The left half showcases the overall testing setup.  The Q-Pix front-end board is positioned inside a shielding box to minimize noise, while the FPGA board is outside.  A waveform generator generates the input voltage, and the CSA output signal and the reset waveform are observed on an oscilloscope.  The lower right part illustrates the Q-Pix front-end board inside the shielding box.  Here, measures such as employing a guarded triple coaxial connector, adding mechanical grooves to the PCB, and removing the solder mask layer above the input section have been adopted to mitigate leakage current.  The upper right section displays the MOSFET, which is directly wire-bonded to the Q-Pix front-end board, serving a key role in the replenishment scheme.}
\end{figure}

The replenishment scheme incorporating the clocked reset represents the implementation for the discrete component form of the Q-Pix front-end, as shown in Fig.~\ref{fig:setup} which displays entire setup.

The critical nature of the MOSFET and the challenge to find a commercially available component with $\mathcal{O}($pA) leakage current encouraged our setup to use an ``in-CMOS external transistor'' for the demonstration circuit.  More importantly, carefully studying the characteristics of this transistor can provide valuable insights for the subsequent Q-Pix ASIC design.  The MOSFET used in this work is a 4-terminal n-channel MOS (NMOS) in ASIC, which is directly wire-bonded to the PCB.  No ESD protection devices are present, which would contribute to sizeable leakage.  The Bulk voltage of the NMOS is adjusted to modify the threshold of MOS and to minimize leakage current through the Bulk-Source junction.

Apart from the MOSFET used for reset, all other components employed are COTS components.  The CSA incorporates an OpAmp LMP7721\cite{LMP7721Datasheet}, which has a low input bias current (typically only \SI{3}{fA}), making it ideal for measuring ultra-low currents in the nanoampere or picoampere range.  Following the CSA is the comparator LT1713\cite{lt1713datasheet} configured as a Schmitt trigger.  %It is worth noting that better decoupling of the comparator is necessary to significantly mitigate the output bouncing of the CSA, as the comparator is part of the large feedback loop.

A guarded triple coaxial connector is used as the input interface to minimize the leakage current of the front-end, and mechanical grooves are added to the PCB without affecting the signal return path.  Additionally, the solder mask layer above the input section is removed during PCB production to reduce the leakage current further.

The required input current signal is generated using a voltage waveform generator in conjunction with on-board current-limiting resistors during testing.  To emulate the current magnitudes originating from LArTPC detectors, a total resistance value of \SI{1}{G\Ohm} was selected.

\section{Transistor characterization for pA-current applications}\label{sec:NMOS}

To design efficient and high-performance Q-Pix electronic circuits for LArTPC, it is crucial to have a comprehensive understanding of NMOS characterization, especially focused on the picoampere range.  In this study, a TSMC \SI{180}{nm} NMOS transistor\cite{TSMC2023}, which has a width ($W$) of \SI{1.2}{\micro\meter} and a length ($L$) of \SI{0.5}{\micro\meter}, was measured to gain insights into its performance characteristics.  Our primary concerns regarding this transistor are whether the drain current (\Ids) can be effectively controlled within the picoampere range, and whether the leakage current remains sufficiently low for the intended application.

\begin{figure}[!htb]
\centering
\includegraphics[width=0.99\linewidth]{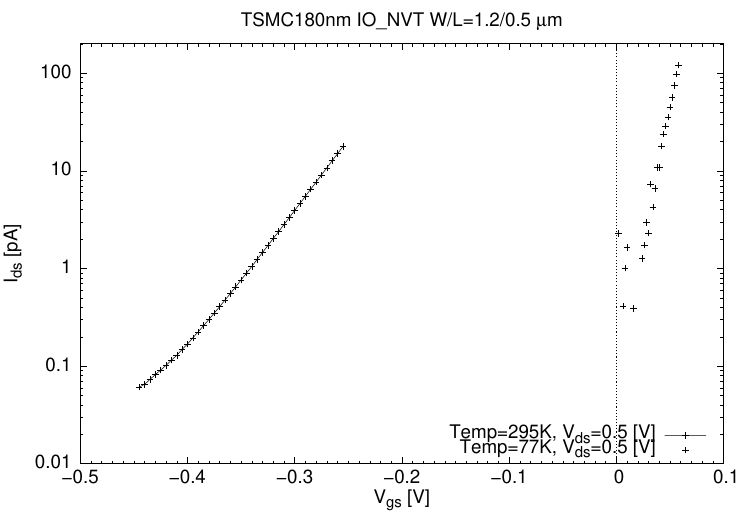}
\caption{\label{fig:ids} $I_{ds}-V_{gs}$ characteristics. This figure demonstrates the relationship between the drain current (\Ids) and the gate-source voltage (\Vgs) for a TSMC \SI{180}{nm} NMOS transistor at two distinct temperatures, \SI{295}{K} and \SI{77}{K}.  It showcases how the transistor can be effectively tuned to operate within the picoampere range, with the need for a lower gate voltage than the source voltage to fully switch off the NMOS.}
\end{figure}

\begin{figure}[!htb]
\centering
\includegraphics[width=0.99\linewidth]{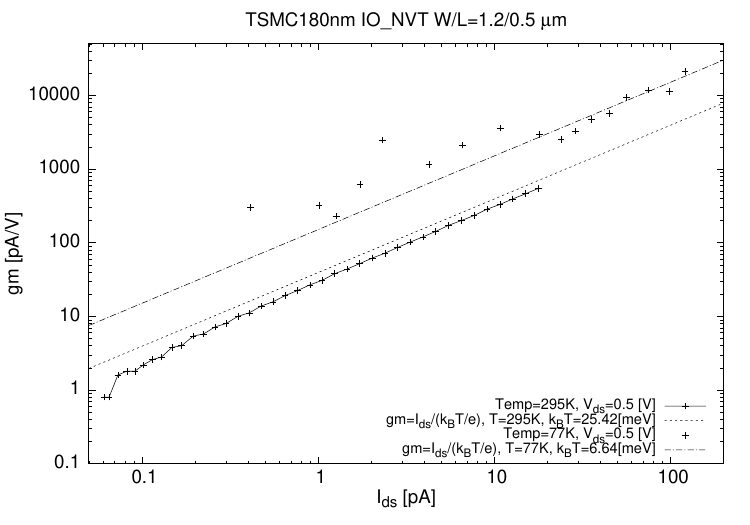}
\caption{\label{fig:gm} Transconductance characteristics.  This figure showcases the transconductance (\gm) results for the NMOS transistor at temperatures of \SI{295}{K} and \SI{77}{K}.  The measured gm values are compared to an ideal curve with a slope factor $n=1$.}
\end{figure}

Fig.~\ref{fig:ids} illustrates the relationship between \Ids and gate-source voltage (\Vgs) at two distinct temperatures, \SI{295}{K} and \SI{77}{K}.  Meanwhile, Fig.~\ref{fig:gm} displays the corresponding transconductance (\gm) results for these conditions.  From Fig.~\ref{fig:ids}, it is evident that the strong inversion region is not appropriate for achieving \si{pA} current levels.  At a temperature of \SI{295}{K}, the transistor exhibits a smooth and consistent transition in the sub-threshold region, reaching currents as low as $\sim$\SI{50}{fA}.  This is attributed to the limitations in the accuracy of the picoammeter employed for measurement.  The observed behavior demonstrates that the drain current can indeed be tuned within the \si{pA} range, and to completely switch off the NMOS, the gate voltage might need to be several hundred millivolts lower than the source voltage.  This voltage level must be maintained for the NMOS in the Q-Pix system for most of its operation time.

For CMOS devices working in the subthreshold region, the transconductance, \gm, is expected to follow
\begin{linenomath*}
    \begin{equation}
% \begin{align}
    g_\mathrm{m} = \frac{I_\mathrm{ds}}{n \cdot k_\mathrm{B} T/e}\,, \label{eq:3.1}
% \end{align}
    \end{equation}
\end{linenomath*}
where $I_{ds}$ is the drain current, $k_B T/e$ is the thermal voltage, and the slope factor $n=(1+C_D/C_{ox})$ is device and process dependent ($C_\mathrm{D}$ and $C_\mathrm{ox}$ are the depletion layer and gate-oxide capacitances, respectively).  For a \SI{180}{nm} process, the typical value of $n$ is 1.4\cite[p.~42-43]{binkley2008}.  This matches the measured data, as shown in Fig.~\ref{fig:gm}.

At \SI{77}{K}, system noise causes some fluctuations in the measured \Ids data.  Despite this noise, the overall trend indicates that the transistor remains capable of regulating the drain current within the \si{pA} range, albeit with less precision compared to \SI{295}{K}.  At \SI{77}{K}, given the same \Vds conditions, the transistor must still operate in the sub-threshold region to control the \Ids at the \si{pA} level, but the required \Vgs is essentially a positive voltage.  This suggests that, in comparison to room temperature, the control circuit for the transistor in liquid argon might not necessitate an additional negative power supply, thereby simplifying the power supply design for Q-Pix.  However, concurrently, the sub-threshold region exhibits a significantly higher sensitivity to gate voltage variations, potentially requiring stricter gate voltage control to maintain the desired pA-level currents.  This increased sensitivity, in turn, places higher demands on the system's noise level.  Though affected by the noise, the \gm values at \SI{77}{K} still show a general adherence to the expected formula.  It is worth noting that \gm is independent of the transistor dimensions.

In conclusion, the experimental results demonstrate that the selected transistor can be effectively tuned to operate in the \si{pA} and \si{nA} range with minimal leakage current, making it well-suited for Q-Pix applications.  The sensitivity of tuning the \Ids in the \si{pA} range is influenced by the transistor's temperature and quality.  To minimize the leakage current when the NMOS is off, the Gate voltage must be maintained below the Source voltage; to achieve nA-level current, the NMOS should operate in the sub-threshold state, allowing for precise control of the current flow in the Q-Pix system.

\section{Heartbeat signals and the time-to-charge linearity}\label{sec:linearity}

A LArTPC detector will have an abiding current due to the \Ar decays.  The average current can be stable over a long period and independent of attachment level.  This radiological background will generate regular ``heartbeat'' reset pulses in a Q-Pix frontend.  Because of ASIC manufacturing variations, individual pixel feedback capacitance and the difference in reset baseline and reset threshold voltages will vary, resulting in significant differences in heartbeat rate among pixels.  This heartbeat can serve as a means to calibrate the charge sensitivity of each pixel automatically.  The ability to perform absolute charge auto-calibration using the intrinsic \Ar decay current is also a major advantage of the Q-Pix readout system.

For validating the functionality of the Q-Pix front-end implementation, the primary concern is whether it can correctly generate reset pulses and if the density or frequency of these pulses is accurately correlated with the input current.  During our test of the Q-Pix front-end board, a simulated ``heartbeat'' can be introduced to calibrate the charge sensitivity by applying a constant voltage to the Q-Pix input through the input current limiting resistors.  If the input voltage is constant, the current will be constant.  This method can also be used to characterize the time-to-charge linearity of the Q-Pix front-end board by changing the input current and measuring the frequency of the reset pulse.

\begin{figure}[!htb]
\centering
\includegraphics[width=0.99\linewidth]{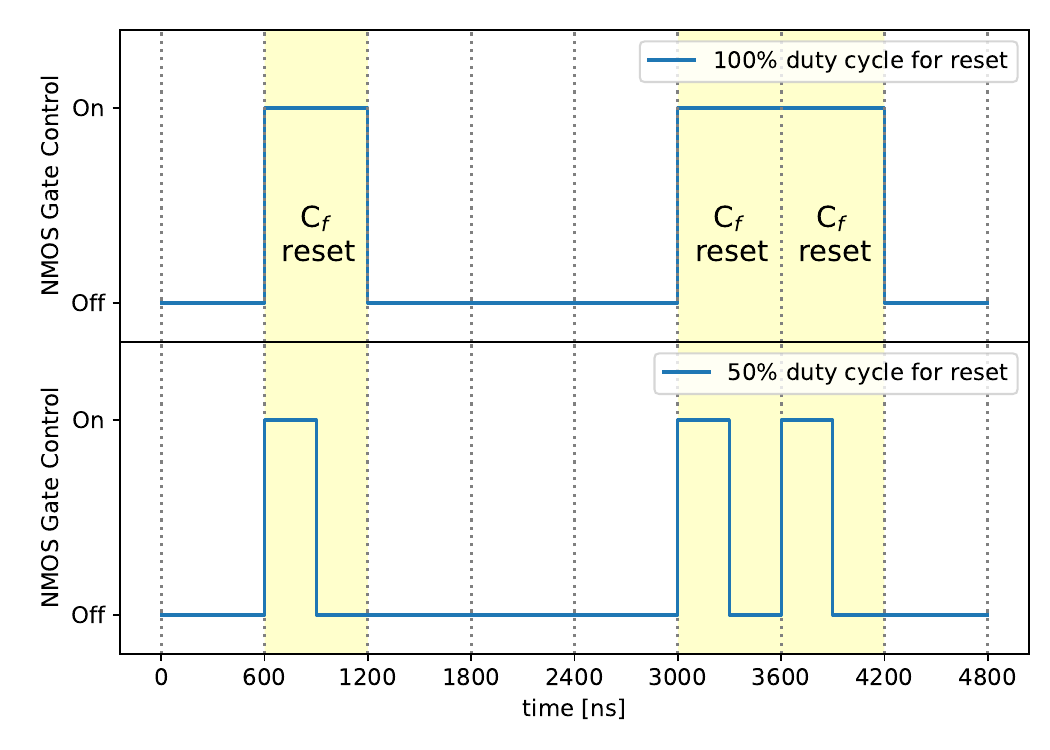}
\caption{\label{fig:resets} Reset pulse shape.  The top panel illustrates a reset pulse with a 100\% duty cycle, while the bottom panel demonstrates a pulse with a 50\% duty cycle.  By ensuring that the duty cycle remains below 100\%, each reset pulse is guaranteed to have both a rising and falling edge, a crucial factor in maintaining a consistent quantization step, \dQ, and enhancing the time-to-charge linearity performance.}
\end{figure}

For achieving good time-to-charge linearity, it is essential to ensure that every quantization step, \dQ, remains consistent.  It is beneficial to maintain a duty cycle for the \Cf reset pulse below $100\%$ and well controlled.  Fig.~\ref{fig:resets} provides two examples with different duty cycles of $100\%$ and $50\%$.  When multiple reset pulses occur in quick succession and are tightly connected, some pulses may miss their positive or negative edges.  Consequently, the \dQ for these pulses may change, depending on the dynamic response of the NMOS to the on and off switching edges.  This variation, in turn, affects the shape of the reconstructed waveform.  When the duty cycle is less than $100\%$, each reset pulse is guaranteed to have both the rising and the falling edges, which helps to maintain a consistent \dQ for each reset.  By tuning the duty cycle, a secondary and digital method can be provided to roughly adjust the NMOS current, apart from controlling the analog voltage of the NMOS gate.  In our application, an equivalent sampling rate of \SI{1.6}{MHz} implies a \SI{600}{ns} reset period, for which a $50\%$ duty cycle was adopted.

\begin{figure}[!htb]
\centering
\includegraphics[width=0.99\linewidth]{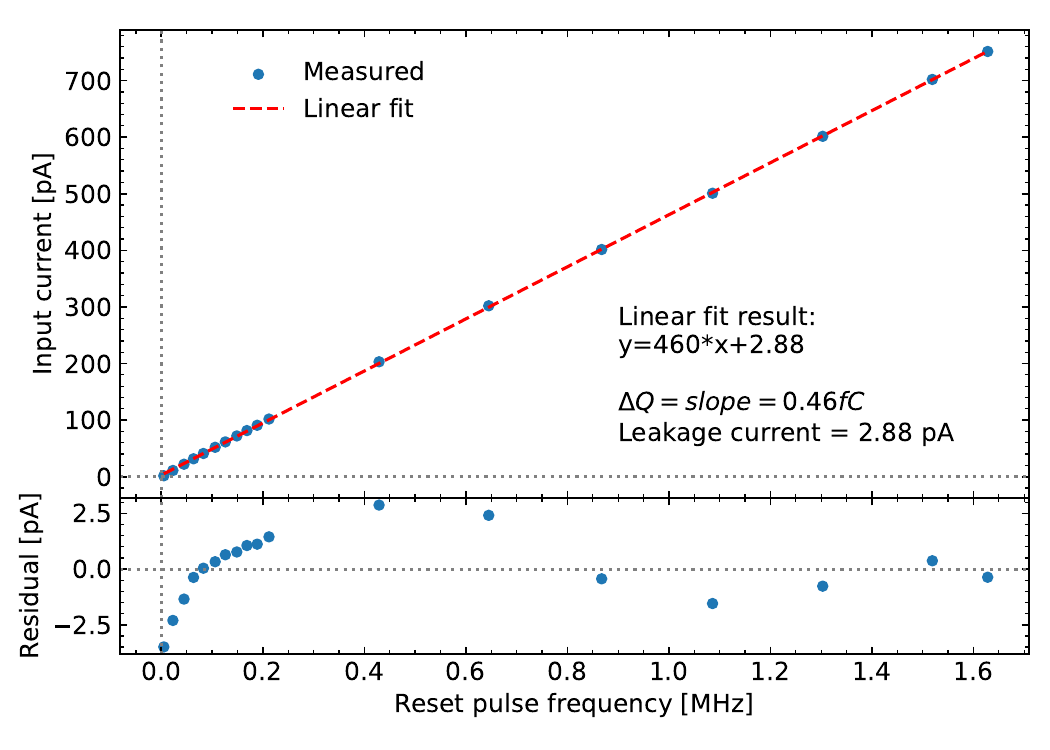}
\caption{\label{fig:linearity} \dQ measurement and the time-to-charge linearity.  This figure demonstrates the relationship between the input current and the reset pulse frequency.  It confirms that \dQ can reach \SI{0.46}{fC} and displays good linearity even down to a few picoamperes.}
\end{figure}

The measured time-to-charge linearity is shown in Fig.~\ref{fig:linearity}.  The relationship between the input current ($I$) and the reset pulse frequency ($f$) can be expressed as
\begin{linenomath*}
    \begin{equation}
% \begin{align}
    I = \Delta Q \cdot f + I_\mathrm{o}\,, \label{eq:4.1}
% \end{align}
    \end{equation}
\end{linenomath*}
where $I_o$ is the leakage current.  In this case, the measurement of our Q-Pix implementation demonstrates that the leakage current can be as small as \SI{3}{pA}, and the \dQ can be \SI{0.46}{fC}, which is already in the expected range.  The figure clearly shows excellent linearity, even down to a few picoamperes.  This result is achieved with a discrete-component design, making a future Q-Pix ASIC design very promising.

\section{Sigma-Delta modulator behavior}\label{sec:sigmaDelta}

\subsection{Q-Pix front-end: a 1st-order Sigma-Delta modulator}

The Q-Pix concept originates from the simple principle of ``Least Action''; however, upon closer inspection, it can be shown that replenishment scheme shares similarities with a more complex and well-established technique, the Sigma-Delta modulator\cite{sigmaDelta,sigma}.

\begin{figure}[!htb]
\centering
\includegraphics[width=0.95\linewidth]{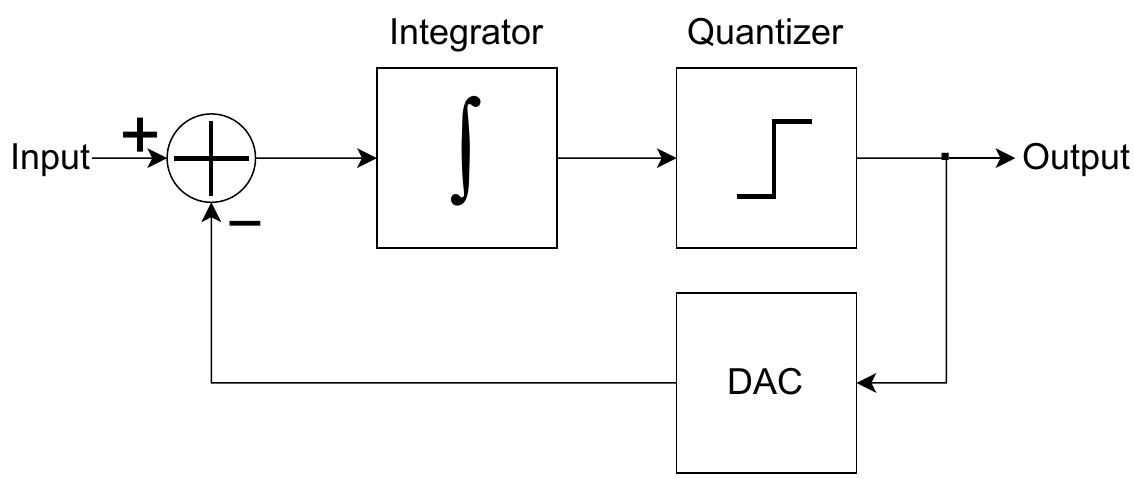}
\caption{\label{fig:sd} Sigma-Delta modulator schematic.  Key components include a quantizer, an integrator, and a low-resolution DAC, which work together to convert the input signal into a high-resolution digital output.}
\end{figure}

A Sigma-Delta modulator (also known as a Delta-Sigma modulator) is an analog-to-digital converter (ADC) with a feedback loop that uses oversampling, noise shaping, and digital filtering to achieve high levels of accuracy and resolution.  As shown in Fig.~\ref{fig:sd}, the feedback loop includes a quantizer, an integrator, and a low-resolution digital-to-analog converter (DAC).  The input signal is subtracted from the DAC output to generate the quantization error signal, which is then filtered by the integrator.  The output of the integrator is then fed into the quantizer, which maps the error signal onto an oversampled digital signal.  After quantization, a low-pass digital filter can be applied to convert this oversampled signal into a high-resolution digital output.

In the case of the Q-Pix, when viewed as a Sigma-Delta modulator, the Schmitt trigger clearly serves as the quantizer, and the reset MOSFET switch essentially functions as a one-bit DAC.  The CSA in the Q-Pix circuit plays a dual role: it amplifies the input charge signals from the TPC detectors and also integrates, or low-pass filters, the quantization error signal generated by the Schmitt trigger and the one-bit DAC.  This filtered signal is then fed to the Schmitt trigger as part of the feedback loop, which helps to reduce the quantization noise generated during the quantization process.

Before viewing the Q-Pix as a Sigma-Delta modulator, one very direct and fundamental quantum of information captured by the Q-Pix was the time difference between one clock capture and the following sequential capture, known as the Reset Time Difference (RTD). This RTD measures the time required to integrate a predefined \dQ.  The arrival of significant signal current leads to very short time intervals between resets on the scale of microseconds.  In other words, input current and reset time differences are inversely correlated: $I \propto 1/RTD$, where $I$ is the average current over an interval $\Delta T$ such that $I\cdot \Delta T=\int I(t)dt=\Delta Q$.  By tracking the sequence of RTDs associated with a particular pixel, back-end computing systems can determine whether a signal is present and extract information about the input current.

The Sigma-Delta-modulator perspective of the Q-Pix design provides a deeper understanding of its operation and introduces new possibilities for information extraction.  The original method of extracting information from the Q-Pix was through RTD measurement; however, this Sigma-Delta-modulator perspective allows for a general low-pass filter to extract input information from the Q-Pix.  This approach permits greater flexibility in the type of information that can be extracted from the Q-Pix and may lead to more efficient data analysis and expanded potential applications.

\subsection{Simulated and measured Sigma-Delta spectral response}

To investigate the behavior of the Q-Pix as a Sigma-Delta modulator, a full-board SPICE\cite{SpiceNagel:M382, engelhardt2015spice} simulation was performed, though not all the parasitic parameters on the PCB could be included. In this simulation, a sinusoidal signal was generated and injected into the Q-Pix front-end board, and the frequency domain response was observed. The simulation was set up to match the specifications of the Q-Pix board, including the sampling frequency, input current, reset pulse duty cycle, and oversampling ratio. By simulating the behavior of the Q-Pix board in this way, it was possible to gain insights into its performance as a sigma-delta modulator.

\begin{figure}[!htb]
\centering
\includegraphics[width=0.99\linewidth]{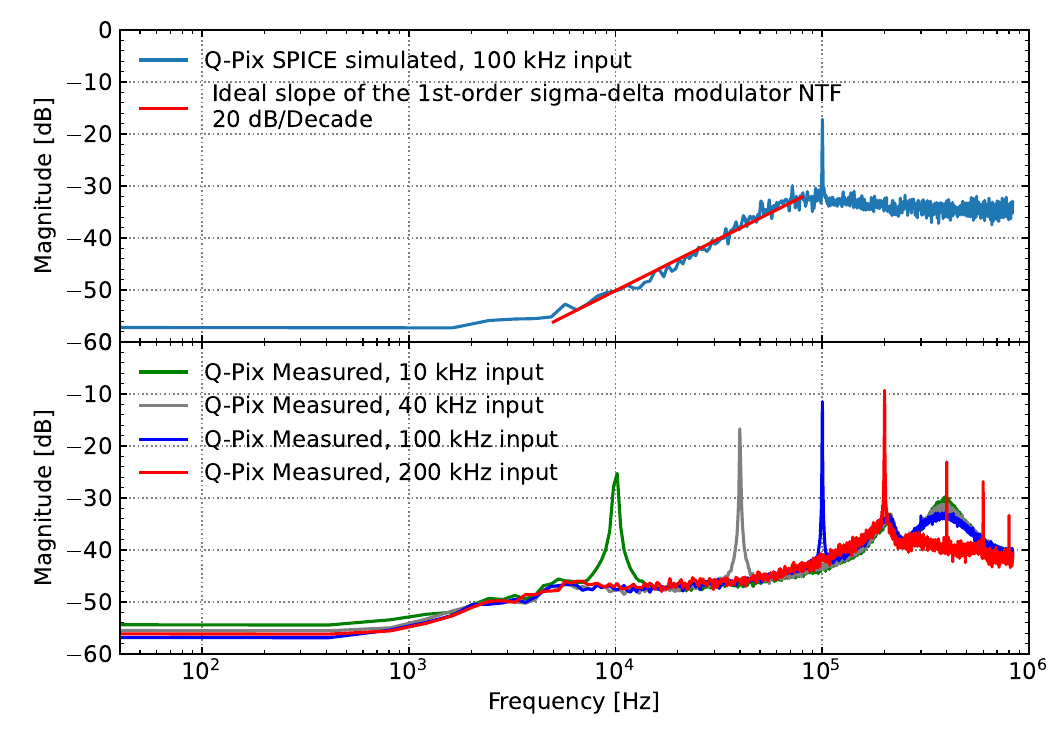}
\caption{\label{fig:sdspec} Simulated and measured Sigma-Delta spectral response. The upper panel illustrates the simulation of a \SI{100}{kHz} sine wave input, confirming the noise-shaping behavior of an ideal Sigma-Delta modulator. The lower panel presents the measured responses, which deviate from the ideal due to complex parasitic parameters that introduce additional low-pass filtering effects. Despite this, the Q-Pix board operates effectively as a first-order Sigma-Delta modulator.}
\end{figure}

The upper panel of Fig.~\ref{fig:sdspec} depicts the simulation results of the Q-Pix board when a \SI{100}{kHz} sine wave input is applied. The noise transfer function (NTF) of an ideal Sigma-Delta modulator suggests that the quantization noise in the low-frequency range is attenuated and shifted to higher frequencies, exhibiting a slope of \SI{20}{dB} per decade---a phenomenon known as noise shaping.  This clearly demonstrates that the full-board simulation results of Q-Pix align well with this noise-shaping behavior.

The lower panel of Fig.~\ref{fig:sdspec} presents the measured results of the Q-Pix front-end board. As can be seen, the noise is indeed pushed to higher frequencies. However, the noise shaping slope is less steep than the \SI{20}{dB} per decade observed in the simulation.  This can be attributed to the actual board containing numerous complex parasitic parameters challenging to model accurately in simulation.  These parasitic parameters result in additional low-pass filtering effects, which lead to a reduced NTF spectral slope when combined with noise shaping.  Nevertheless, based on the test results with input sine waves of different frequencies, the impact of this low-pass effect on the signal can be considered negligible.  Thus, it helps improve the system's signal-to-noise ratio of our Q-Pix implementation.

In any case, the Q-Pix replenishment scheme is indeed a first-order Sigma-Delta modulator.  This implies that a general digital low-pass filter can still be used to enable our PCB-scale design of the Q-Pix to achieve better resolution and waveform reconstruction. Furthermore, this measurement provides a valuable reference for understanding how parasitic parameters in a PCB-scale discrete-component system can affect the Q-Pix's behavior in such low-current and sensitive applications. In an ASIC-scale design, it can be expected that the performance will be significantly improved due to reduced parasitics and enhanced integration.

\section{Waveform reconstruction: comparing simulation and measurements}\label{sec:reconstruction}
\subsection{First demonstration of the real Q-Pix waveforms: simulated vs. measured}

\begin{figure}[!htb]
\centering
\includegraphics[width=0.99\linewidth]{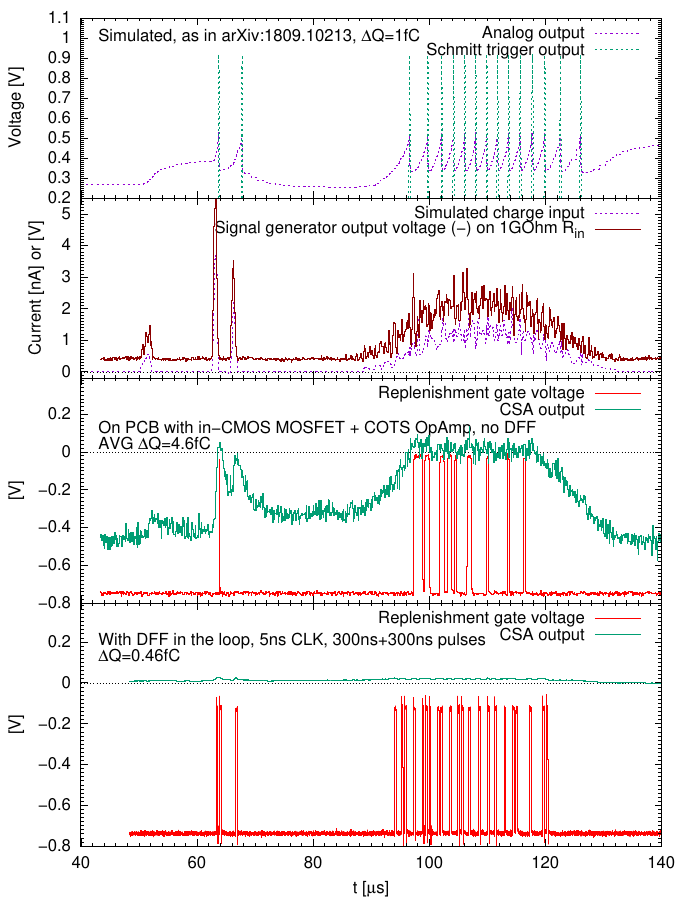}
\caption{\label{fig:resetWaves} Q-Pix output waveforms: simulated vs measured.  The top two panels display the simulated transistor-level Charge Integration results for a typical Min-I track current in LAr and the input waveform.  While the baseline of both curves in the second panel is indeed \SI{0}{V}, and the amplitudes are the same, the output waveform from the Signal Generator has been intentionally shifted upward for visual clarity---the bottom two panels present measured outputs from the Q-Pix front-end board.  The 3rd panel represents the initial setup without a DFF, clocked reset, or proper shielding, resulting in noise interference and inconsistent \dQ.  In the 4th panel, after implementing a clocked reset and improving the setup, a significantly reduced \dQ of \SI{0.46}{fC} is achieved, demonstrating a promising correlation between the reset waveform and the input current.}
\end{figure}

Q-Pix is designed to achieve precise measurement and high-quality capture of spatial and topological information, energy loss, and event time for a true signal event.  A conventional waveform of the input signal from LArTPC detectors can be directly reconstructed by converting time-stamped charge into current.  The top panel of Fig.~\ref{fig:resetWaves} shows the simulated output results of transistor-level Charge Integration simulation for a generic Min-I track current in LAr.  The input waveform of this generic Min-I track current is also displayed in the second top panel.  The simulation was performed with a \dQ of \SI{1}{fC}, following the basic reset scheme as shown in Fig.~\ref{fig:schemes}(a).  The sharp pulses in the top panel represent the Schmitt trigger output (resets) that can be used for waveform reconstruction after being time-stamped.

To measure the Q-Pix front-end board, the simulated in-LAr charge current waveform is first fed into a signal generator to produce a voltage waveform.  This is then applied to a series of resistors with a total resistance of \SI{1}{G\Ohm} to create an appropriate input current, as shown in the second panel of Fig.~\ref{fig:resetWaves}.  Our primary focus is on the resulting output waveform from the Q-Pix front-end board and whether it can correctly output reset pulses correlated with the input current signal shape.

The 3rd and 4th panels of Fig.~\ref{fig:resetWaves} present two measured output waveform results of our Q-Pix implementation in the replenishment scheme. The green curve represents the output waveform of the CSA, while the red pulses correspond to the Q-Pix reset output.

The 3rd panel corresponds to a test for the Q-Pix implementation before improvements, without using a DFF or clocked reset, or proper shielding.  Due to the lack of adequate shielding, noticeable noise can be seen superimposed on the green CSA output waveform. As explained in section \ref{sec:design}, the absence of a clocked reset, noise interference, and insufficient dynamic response speed of the CSA will result in an inability to maintain a consistent \dQ throughout the measurements.  This is reflected in the randomness of the pulse width in the red reset waveform.  At the same time, this also limits the achievable minimum \dQ.  The average \dQ of \SI{4.6}{fC} in the 3rd panel is already quite close to its minimum limitation, but it is clearly too large for the desired range of \SI{0.3}{fC} to \SI{1}{fC} that we aim to achieve.

In contrast, in the 4th panel, after implementing the clocked reset with an appropriate duty cycle, we successfully achieved a \dQ of only \SI{0.46}{fC}, which is only a tenth of the original \dQ in the 3rd panel.  The green waveform, representing the CSA output, serves as the error signal for the Sigma-Delta modulator.  The amplitude of this signal is minimal, and more detailed attributes of this waveform will be further displayed in Fig.~\ref{fig:reconstruction}.  While \SI{0.46}{fC} does not represent the lowest threshold this implementation of the frontend could achieve, further reduction in \dQ was not the primary aim of this study.  It is worth noting that this demonstration already exceeds the performance requirement for Q-Pix and thus builds confidence of the ability to achieve this level in a silicon implementation of Q-Pix.

Both measurement results in the 3rd and 4th panels demonstrate the correct correlation between the reset waveform and the input current waveform.  As observed, when there is a large amount of charge input, the reset pulses become more dense, consistent with the scenario described by the RTD.  This is precisely what we expect, and it is the first demonstration of the real Q-Pix waveforms.  The remaining question becomes: How best to reconstruct the waveform of the input current?

\subsection{Current waveform reconstruction: simulated vs. measured}

Once Q-Pix is treated as a Sigma-Delta modulator, a low-pass filter can be used directly for waveform reconstruction based on reset outputs.  In this case, a convenient choice for the low-pass filter is the Cascaded Integrator-Comb (CIC) filter\cite{CIC1981,CIC2022}.  The CIC filter is a linear, finite impulse response (FIR) filter widely used in digital signal processing.  The main advantage of the CIC filter is its simplicity, as it does not require any multipliers making it computationally efficient and suitable for hardware implementation.  The CIC relies on a combination of integrators and comb filters composed of adders, subtractors, and delays.  CIC filters are particularly well-suited for applications with constant input and output sampling rates, and they perform optimally when the ratio of these rates is an integer value.

In the context of Q-Pix, the CIC filter can be employed for waveform reconstruction based on reset outputs.  Owing to its simplicity and computational efficiency, real-time waveform reconstruction using CIC filters in hardware can be performed if necessary. Moreover, CIC filters can easily be cascaded, meaning multiple stages can be combined to achieve the desired filter response with increased precision and complexity.  This makes it an attractive choice for processing the output of the Q-Pix front-end in a LArTPC detector system.

\begin{figure*}[!htb]
\centering
\includegraphics[width=0.95\linewidth]{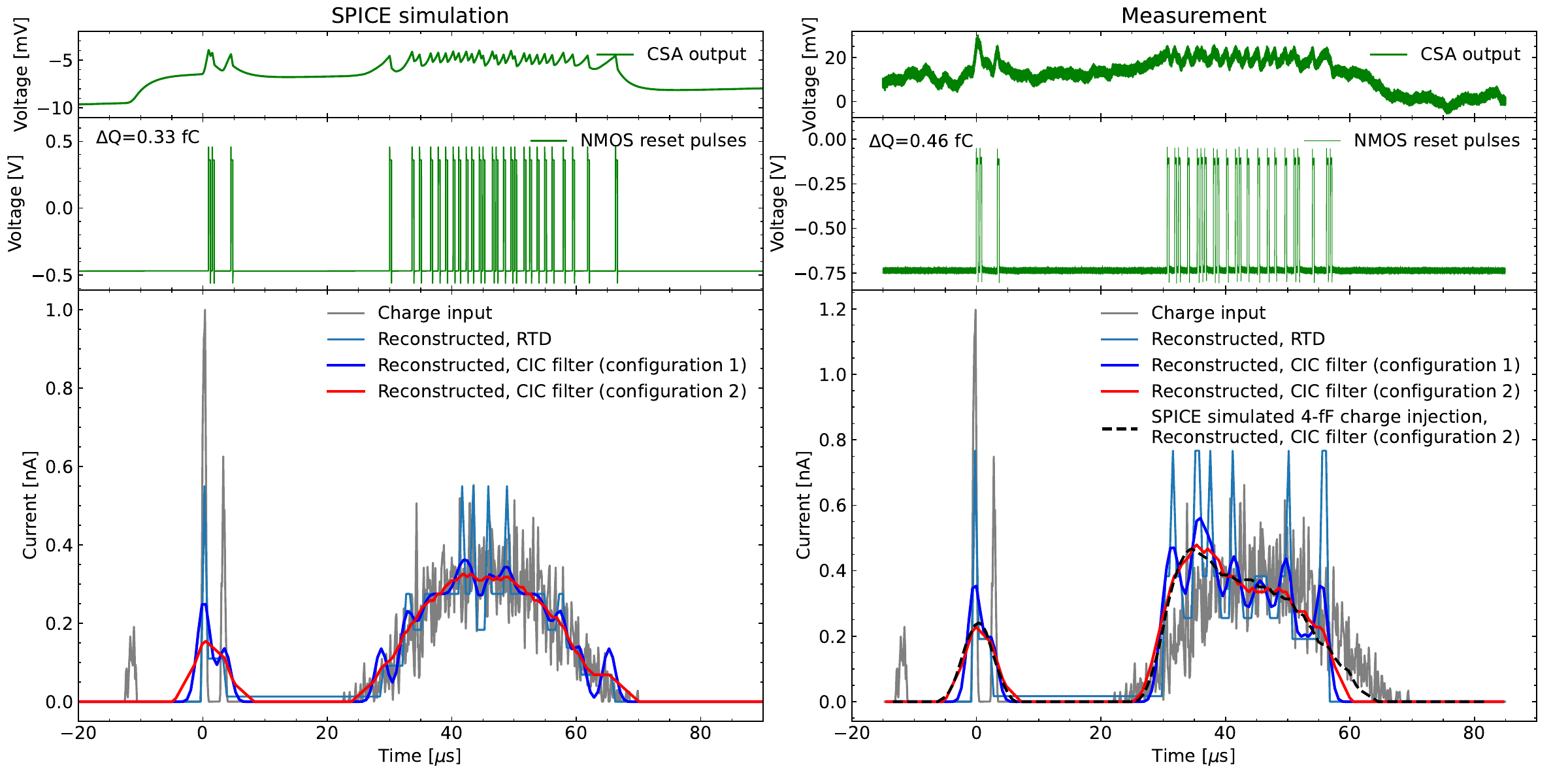}
\caption{\label{fig:reconstruction} Current waveform reconstruction: simulated vs. measured.  This figure showcases the application of CIC filters in reconstructing Q-Pix output waveforms from SPICE simulations and actual measurements.  In the SPICE simulations, the waveforms reconstructed with two different CIC filter bandwidths closely match the input signals, affirming the viability of digital low-pass filters in waveform reconstruction.  The left side shows the waveform reconstructed from SPICE simulations based on the actual PCB design of the Q-Pix front-end board.  On the right side, the figure presents waveform reconstruction results from actual measurements, highlighting some shape differences due to the charge injection effect, primarily originating from the package parasitic capacitance of the input path resistor. However, after introducing a \SI{4}{fF} parasitic capacitance in the SPICE simulation, the resulting reconstructed waveform closely matches the actual measurement, further validating the effectiveness of Q-Pix in reconstructing waveforms despite external charge injection effects.}
\end{figure*}

In our study, we first focused on applying CIC filtering to the output waveforms of Q-Pix SPICE simulations.  This allows us to validate the feasibility of waveform reconstruction using digital low-pass filters.  The left side of Fig.~\ref{fig:reconstruction} displays the results of waveform reconstruction obtained from the SPICE simulations.  It is worth noting that these simulations differ from those presented in Fig.~\ref{fig:resetWaves}.  Fig.~\ref{fig:resetWaves} represents the principle simulation for the CIR structure of Q-Pix, while Fig.~\ref{fig:reconstruction} showcases the SPICE simulation results based on the actual PCB design of the Q-Pix front-end board.

The bottom-left panel in Fig.~\ref{fig:reconstruction} presents two waveforms reconstructed with two different CIC filters, each characterized by a unique combination of first zero frequency and number of stages.  The blue curve corresponds to a CIC filter with its first zero at approximately \SI{800}{kHz} and is composed of 10 stages.  The red curve is a CIC filter with the first zero at approximately \SI{200}{kHz} and is composed of 2 stages.  The reconstructed waveforms  using directly the RTD's as well as the two distinct CIC filters all closely align with the input signals.

Compared to RTD, digital filters such as CIC offer more flexibility in reconstruction.  Generally speaking, a filter with a lower bandwidth offers more effective noise removal, resulting in smoother reconstructed waveforms.  In contrast, a filter with a higher bandwidth may provide better time resolution but at the cost of retaining more noise.  This level of flexibility facilitates the optimal calibration between noise reduction and time resolution, enabling adjustments tailored to the distinct needs of the physics application.  These SPICE simulation results demonstrate that Q-Pix can successfully reconstruct input charge waveforms using generic digital low-pass filters.

Some unexpected issues arise when we shift our attention to the waveform reconstruction results from measurements using our demonstrator board.  The waveform reconstruction results from the Q-Pix front-end board are shown on the right side of Fig.~\ref{fig:reconstruction}, using the same CIC filter configurations as in the simulations.

The total charge amount after waveform reconstruction is verified as equal to the input charge.  However, as seen in the figure, there are noticeable shape differences between the reconstructed waveforms and the input waveforms, primarily manifesting as the charge appearing to be unexpectedly concentrated near the rising edge of the input current waveform.

This phenomenon was discovered to be caused by a charge injection effect.  This effect comes from the package parasitic capacitance of the resistor in the input path.  In the subsequent section, we aim to quantify the magnitude of this parasitic capacitance for the resistor deployed

\subsection{Charge injection effect}

\begin{figure}[!htb]
\centering
\includegraphics[width=0.99\linewidth]{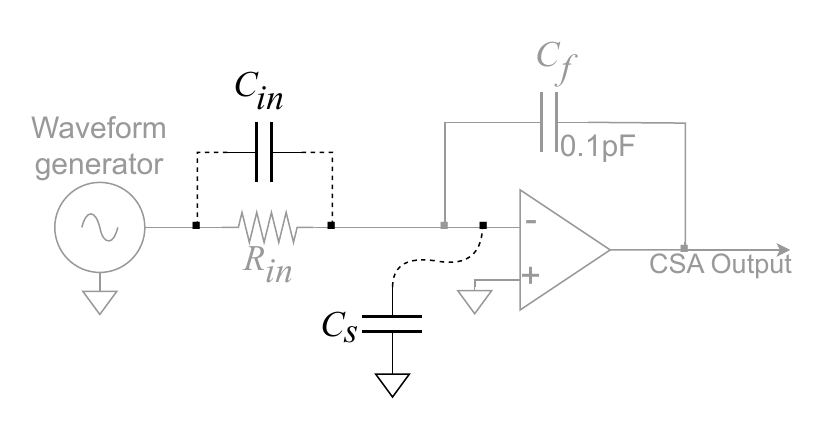}
\caption{\label{fig:Cin} Key parasitic capacitances. This figure illustrates the locations of two significant parasitic capacitances, \Cs and \Cin that substantially impact the Q-Pix's input-output behavior. Stray capacitance \Cs primarily influences the CSA's response speed and significantly affects the reset scheme's performance.  Meanwhile, \Cin is associated with the packaging of the input resistor \Rin, and it plays a role in the input charge.}
\end{figure}

The first task is to locate where in the circuit these parasitic capacitances arise.  When the feedback capacitance \Cf of Q-Pix is already on the same order of magnitude as the parasitic capacitances, the ubiquitous parasitic capacitances across the entire PCB all become suspect.  The key is to identify which parasitic capacitance significantly impacts the performance.  Through an iterative process of conjecture, simulation, elimination, or verification, we ultimately conclude that there are two distinct parasitic capacitances.  These are stray capacitance (\Cs) and parasitic capacitance accompanying the input resistor (\Cin).  These are shown schematically in Fig.~\ref{fig:Cin}.

\Cs is the total stray capacitance at the inverting input of the CSA, which originates from the amplifier input capacitance, and any other stray capacitance from the circuit board traces.  The first part can be directly obtained through SPICE simulation.  Every amplifier has some capacitance between each input and AC ground, and typically, this input capacitance is one of the critical parameters of the SPICE simulation model.  According to the LMP7721 datasheet, the typical value of this amplifier's input capacitance is \SI{11}{pF}.  Based on the SPICE simulation results in the Q-Pix application, the total \Cs can be calculated as \SI{20}{pF}, which is close to this value.  In our example, it is worth noting that \Cs is approximately \si{200} times larger than \Cf.  Furthermore, \Cs is located directly at the input of the amplifier, which can result in a reduced response speed for the CSA.

The positive aspect is that, apart from reducing the response speed, \Cs does not impact the functionality of the replenishment scheme. In the replenishment scheme, the value of \dQ is determined by the product of the NMOS current and the reset pulse width, which is independent of \Cs. According to the SPICE simulations of the replenishment scheme implementation, each reset-injected \dQ will be temporarily ``buffered'' on \Cs during the brief period when the NMOS is turned on.  After the NMOS is turned off and the CSA settles, all the charge initially buffered on \Cs will be redistributed and correctly transferred to \Cf. Consequently, \Cs does not affect the \dQ or the charge measurement of the replenishment scheme. Although \Cs impacts the system bandwidth or the maximum equivalent sampling rate, an equivalent sampling rate of \SI{1.6}{MHz}, as previously mentioned, is already sufficient for our Q-Pix demonstration.

However, the situation changes for the reset scheme, as simulation results indicate that a large \Cs can significantly impact its performance. The reset scheme relies on rapidly turning on the MOSFET to discharge \Cf when its voltage reaches the threshold, with \dQ being the amount of charge discharged from \Cf for each reset. Ideally, the time constant \CfRds limits the minimum reset width, where \Rds represents the resistance between the source and drain of the MOSFET when it is turned on. But when \Cs is present and cannot be neglected, turning on the MOSFET to discharge \Cf also causes the CSA output to charge \Cs through the MOSFET. As \Cs is much larger than \Cf, the unknown charge in \Cs after charging significantly affects \dQ. Consequently, the time constant \CsRds, rather than \CfRds, limits the practically achievable minimum reset width, which in our case would be \si{200} times the original value. According to the simulation, the \Rds of the chosen NMOS is approximately \SI{40}{k\Ohm}, and \Cs is \SI{20}{pF}. Assuming a reset width of $2\CsRds = \SI{1600}{ns}$ is selected, the maximum equivalent sampling rate is limited to \SI{625}{kHz}.  Moreover, an excessive reset width may lead to a significant input charge loss in the reset scheme; for an input current of \SI{1}{nA}, a single reset pulse could result in a \SI{1.6}{fC} input charge loss.

Therefore, a large \Cs suggests that, in principle, our discrete component solution cannot implement the reset scheme.  While this is present in the discrete component implementation of the Q-Pix frontend, this should not be present in a fully silicon ASIC design in which there is more critical control over these aspects.

\begin{figure}[!htb]
\centering
\includegraphics[width=0.99\linewidth]{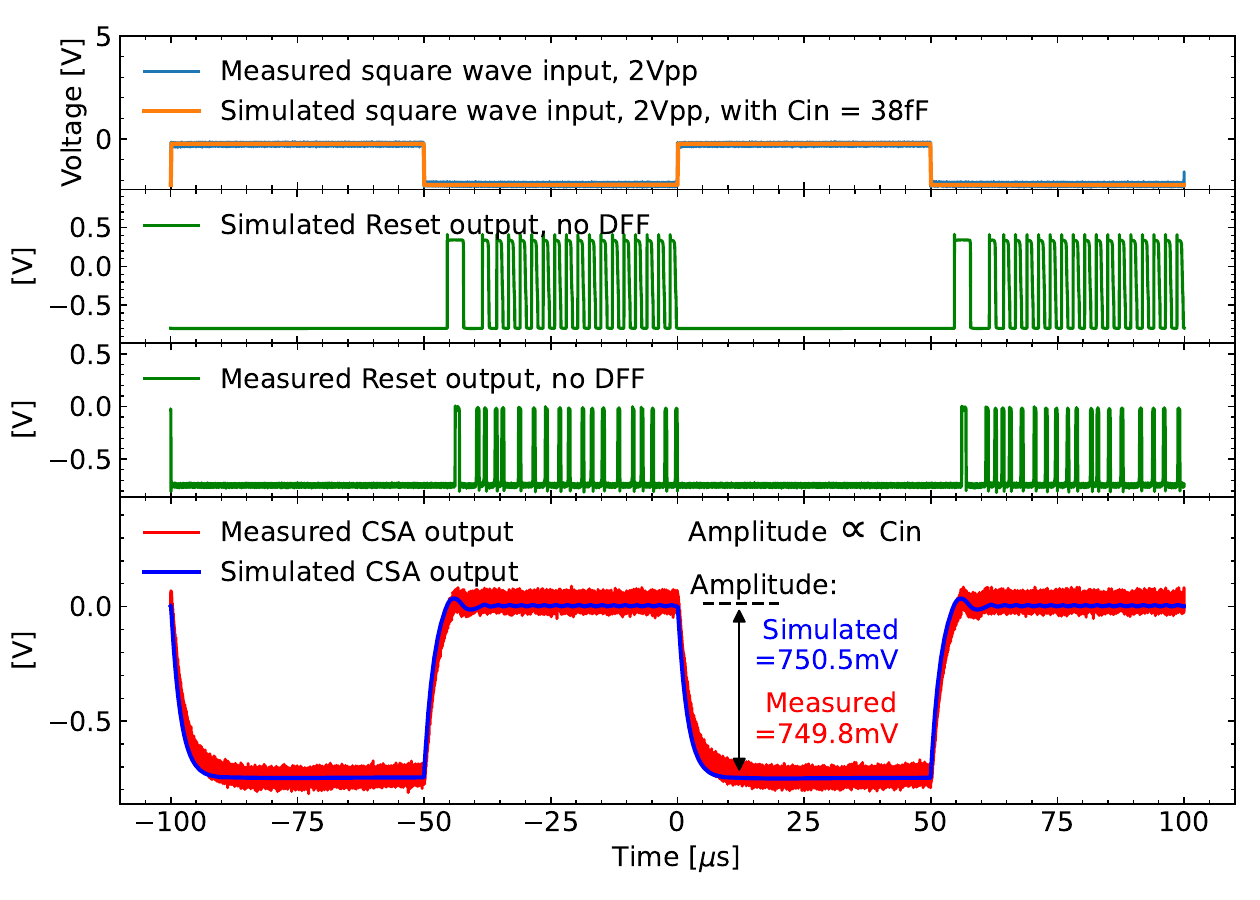}
\caption{\label{fig:cInj} Charge injection: simulated vs measured. This figure presents the measurement and simulation results of parasitic capacitance \Cin. Using a square wave input to induce charge injection, \Cin's measurement is inferred from the amplitude of the CSA output waveform. The figure shows a strong correlation between the simulated and actual measurements, validating the effects of \Cin on the Q-Pix front-end board.}
\end{figure}

\Cin primarily originates from the parasitic capacitance associated with the package of the input resistor \Rin. Although effectively in parallel with \Rin, it is in series in the input path of Q-Pix. \Cin can be measured directly using a method that involves applying a square wave input. At the edge of the square wave, charge injection occurs, causing an inverse voltage drop at the CSA output relative to the input. The waveform amplitude of the CSA output ($\Delta\Vout$) is proportional to \Cin, allowing \Cin to be expressed as:
\begin{linenomath*}
    \begin{equation}
% \begin{align}
    C_\mathrm{in}=\frac{Q_\mathrm{injected}}{\Delta V_\mathrm{in}}=\frac{C_\mathrm{f}}{\Delta V_\mathrm{in}}\cdot {\Delta V_\mathrm{out}}\,,\label{eq:6.1}
% \end{align}
    \end{equation}
\end{linenomath*}
where \D \Vin is the amplitude of the input square wave. Fig.~\ref{fig:cInj} displays the measurement results of \Cin. With $\Delta \Vin=\SI{2}{Vpp}$ and a measured output amplitude of \SI{750.5}{mV}, \Cin is determined to be \SI{38}{fF}. This represents the parasitic input capacitance caused by the package of a single \SI{1}{G\Ohm} resistor with dimensions of \SI{3.2}{\milli\meter} $\times$ \SI{1.6}{\milli\meter}.

In SPICE simulations, \Cs is already included in the parameters of the amplifier model. However, since \Rin is an ideal resistor, \Cin does not exist. To simulate \Cin, an appropriate parallel capacitance must be deliberately added, allowing the influence of \Cin on the Q-Pix front-end board to be observed through SPICE simulations. After incorporating a \SI{38}{fF} parasitic capacitance, the simulated CSA output amplitude is \SI{749.8}{mV}, with the output waveform also presented in Fig.~\ref{fig:cInj}. The simulation result closely aligns with the actual measurements.

By replacing the original \SI{1}{G\Ohm} resistor with an arrangement of ten \SI{100}{M\Ohm} resistors connected in series, the total parasitic capacitance \Cin can be significantly reduced to approximately one-tenth of its initial value, or around \SI{4}{fF}. The measurement setup in Fig.~\ref{fig:reconstruction} already adopts this optimized configuration.

To verify that the observed difference in the reconstructed waveform shape is indeed due to \Cin, a \SI{4}{fF} capacitor was added in parallel to the input current limiting resistor in the SPICE simulation. The simulation results show that the charge injection effect causes reset pulses to become more focused at the front edge of the input waveform, although it doesn't impact the total charge measurement.

When the same CIC filter is applied to the simulation-generated reset pulse for waveform reconstruction, a deviation in the waveform shape becomes apparent, as illustrated by the black dashed curve on the lower right of Fig.~\ref{fig:reconstruction}.  After taking into account the charge injection from the \SI{4}{fF} \Cin, this simulated reconstructed waveform aligns closely with the actual measurement, which is depicted by the red reconstructed waveform.

This conclusively demonstrates that the actual measurements validate Q-Pix's ability to reconstruct waveforms of the input current effectively.  The observed deviation in the reconstructed waveform shape arises from external charge injection rather than being an inherent issue with Q-Pix.

\subsection{Summary}

Overall, both the simulations and the measurements compellingly demonstrate Q-Pix's ability to reconstruct \waveforms.  Furthermore, we identified two significant parasitic capacitances, \Cs and \Cin, which could potentially affect Q-Pix's performance.  \Cs primarily impacts the reset scheme, whereas \Cin influences the shape of the reconstructed waveform.

Fortunately, in the final LArTPC application, \Cin will not be present, as there will be no input current limiting resistor for Q-Pix in that configuration. This absence will enhance accuracy and precision in the reconstructed waveform by eliminating the charge injection effects.

\section{Conclusions}\label{sec:summary}

This paper successfully demonstrated the Q-Pix front-end architecture, utilizing COTS OpAmps and in-CMOS external transistors.

The Q-Pix scheme, grounded in the principle of Least Action, presents a promising and innovative approach to charge collection and processing in large-scale LArTPC detectors. Characterization of transistors for pA-current applications offers valuable insights into their performance, establishing their suitability for low-current applications such as the Q-Pix system. A significant advantage of the Q-Pix system is its absolute charge auto-calibration capability, leveraging the intrinsic \Ar decay current.

Examining the Q-Pix system through the lens of a Sigma-Delta modulator allows us to gain a deeper understanding of its operation and uncover new opportunities for information extraction, potentially leading to more efficient or flexible data analysis and an expanded range of applications. Furthermore, the waveform reconstruction results, derived from both SPICE simulations and measurements using the Q-Pix front-end board, showcase promising performance and highlight the potential for precise signal recovery.

The optimistic and encouraging results obtained in this study provide valuable experience and insights for future Q-Pix ASIC design.  With continued research and development, the Q-Pix system has the potential to become a key enabling technology for next-generation detectors in neutrino physics and other high-energy physics experiments.

\section*{Acknowledgments}

This work is supported, in part, by the Laboratory Directed Research and Development (LDRD) funding from Berkeley Lab, provided by the Director, Office of Science, of the U.S.\ Department of Energy under Contract No.~DE-AC02-05CH11231, by the U.S. Department of Energy, Office of Science, Office of High Energy Physics Award No.~DE-0000253485 and No.~DE-SC0020065.  This publication is funded by the Gordon and Betty Moore Foundation through Grant GBMF11565 and grant DOI \url{https://doi.org/10.37807/GBMF11565}.

%\section*{References}

% The \nocite command causes all entries in a bibliography to be printed out
% whether or not they are actually referenced in the text. This is appropriate
% for the sample file to show the different styles of references, but authors
% most likely will not want to use it.
%\nocite{*}

\bibliographystyle{elsarticle-num-names}
\bibliography{refs}% Produces the bibliography via BibTeX.

\end{document}